\documentclass[twocolumn]{aastex61}
\usepackage[T2A]{fontenc}
\usepackage[utf8]{inputenc}
\usepackage[russian,english]{babel}
\usepackage{amsmath}
\usepackage{amssymb} % for \varkappa
\usepackage{upgreek}
\usepackage{enumerate}

\submitjournal{ApJ}

\begin{document}

\title{Weibel instability in hot plasma flows with production of gamma-rays and electron-positron
pairs}

\author{E.~N.~Nerush}
\affiliation{Institute of Applied Physics of the Russian Academy of
Sciences, 46 Ulyanov St., Nizhny Novgorod 603950, Russia}

\author{D.~A.~Serebryakov}
\affiliation{Institute of Applied Physics of the Russian Academy of
Sciences, 46 Ulyanov St., Nizhny Novgorod 603950, Russia}

\author{I.~Yu.~Kostyukov}
\affiliation{Institute of Applied Physics of the Russian Academy of
Sciences, 46 Ulyanov St., Nizhny Novgorod 603950, Russia}

\correspondingauthor{E.~N.~Nerush}
\email{nerush@appl.sci-nnov.ru}

\keywords{gamma-ray burst: general --- instabilities --- methods: numerical --- shock waves}

\begin{abstract}
    We present the results of theoretical analysis and numerical simulations of the Weibel instability in two
    counter-streaming hot relativistic plasma flows, e.g. flows of electron-proton plasma having rest-mass density $\rho \sim
    10^{-4}\; \text{g}\, \text{cm}^{-3}$, Lorentz factors $\Gamma \sim 10$
    and proper temperature $T \sim 10^{13}\; \text{K}$. The instability growth rate and the filament
    size at the linear stage are found analytically, and are in qualitative agreement with
    results of three-dimensional particle-in-cell simulations. In the simulations,
    incoherent synchrotron emission and pair photoproduction in electromagnetic fields are taken
    into account. If the plasma flows are dense, fast and/or hot enough, the overall energy of
    synchrotron
    photons can be much larger than the energy of generated electromagnetic fields. Furthermore, a sizable number of
    positrons can be produced due to the pair photoproduction in the generated magnetic field. We propose a
    rough criterion for judging copious pair
    production and synchrotron losses. By means of
    this criterion we conclude that incoherent synchrotron emission and pair production during
    the Weibel instability can have implications for the collapsar model of gamma-ray bursts.
\end{abstract}

%\maketitle

\section{Introduction}
\label{intro}

The Weibel instability~\citep{Weibel59} is thought to be a source of near-equipartition magnetic field
and a power-law high-energy tail in electron spectra~\citep{Silva03, Saito04, Spitkovsky08a, Nishikawa09} in a plenty of
astrophysical objects, e.g. in gamma-ray bursts (GRBs). The magnetic field lives for a long time due to
nonlinear growth of the field scale~\citep{Silva03, Medvedev05} or even longer due to continuous
particle injection~\citep{Garasev16a}, and ensures prolonged synchrotron emission needed for GRBs
afterglow interpretation~\citep{Piran99}. Synchrotron afterglow model explains the GRB emission fairly well at least in radio
band~\citep{Chevalier98, Soderberg10}. The Weibel instability have been intensively studied
theoretically~\citep{Grassi17}, numerically (including extreme laser fields, see \citet{Efimenko17}) and
experimentally~\citep{Liu11, Huntington15, Garasev17b}.

One may notice that the power of the synchrotron emission is proportional to~\citep{LandauII}
$\gamma^2 B^2$,
where $\gamma$ is the electron Lorentz factor and $B$ is the magnitude of the large-scale
electromagnetic fields. Thus, this power is approximately proportional to the cube of the energy
density of the flows~\cite{Piran99}, and the energy being carried away by synchrotron
photons can become greater than
the energy of large-scale electromagnetic fields for quite dense and energetic flows.
More precisely, we consider plasmas and
fields such that $\chi \sim 1$, where $\chi$ is the quantum parameter crucial for synchrotron
emission~\citep{Berestetskii82}:
\begin{equation}
    \label{chi}
    \chi = \frac{e \hbar}{m^3 c^4} \sqrt{(\varepsilon E / c +
    \mathbf p_e \times \mathbf B)^2 - (\mathbf p_e \cdot \mathbf E)^2},
\end{equation}
where $\varepsilon$ and $\mathbf p_e$ are the electron energy and momentum,
$\mathbf E$ and $\mathbf B$ are the electric and magnetic field magnitudes,
$\hbar$ is the Planck's constant, $c$ is the speed of light, $e>0$ and $m$ are
the electron charge and mass, respectively. If $\chi \gtrsim 1$, the energy of a photon emitted by
an electron is about the electron energy, and the average distance on which the photon emission
occurs is about $\ell_{em} \sim \ell_f / \alpha$, where $\ell_f \sim mc^2 / (eB)$ is the radiation formation
length~\citep{Berestetskii82} and $\alpha = e^2 / \hbar c \approx 1 / 137$ is the fine structure
constant. Hence the ratio of $\ell_{em} / c$ to the timescale of the Weibel
instability~\citep{Grassi17} is the
following:
\begin{equation}
    \label{ell_em}
    \frac{\ell_{em} \omega}{c \bar \gamma_e^{1/2}} \sim \frac{1}{\alpha \bar \gamma_e},
\end{equation}
where $\omega = (4 \pi e^2 n_e / m)^{1/2}$ is the electron plasma frequency, and we use the
equipartition assumption $B^2 \sim 8 \pi n_e m c^2 \bar \gamma_e$, $n_e$ is the electron density,
$\bar \gamma_e$ is the mean electron Lorentz factor. Eq.~(\ref{ell_em})
obviously means
that if $\bar \gamma_e \gtrsim 137$ and $\chi \sim 1$ is reached, the synchrotron emission
potentially can take away the electron energy in a timescale lower than the Weibel instability
timescale. Thus, synchrotron losses should be taken into account if one considers the Weibel
instability in dense ultrarelativistic plasma flows.

If for an electron $\chi \sim 1$, it quite probably emits a photon with momentum $p_\gamma \sim p_e$ almost parallel to the electron
momentum, $\mathbf p_\gamma \parallel \mathbf p_e$ and with the energy about the electron
energy~\citep{Baier98, Berestetskii82},
$\varepsilon_\gamma \sim \varepsilon$.
Pair photoproduction in strong electromagnetic field:
\begin{equation}
    \label{BreitWheeler}
    \gamma \rightarrow e^+ + e^-
\end{equation}
is governed by the quantum parameter
\begin{equation}
    \label{varkappa}
    \varkappa = \frac{e \hbar}{m^3 c^4} \sqrt{(\varepsilon_\gamma E / c +
    \mathbf p_\gamma \times \mathbf B)^2 - (\mathbf p_\gamma \cdot \mathbf E)^2},
\end{equation}
that is the same as $\chi$ (Eq.~(\ref{chi})) with $\varepsilon_\gamma$ and $\mathbf p_\gamma$
substituted for $\varepsilon$ and $\mathbf p_e$, respectively. Hence, for the photon emitted by
electron with $\chi \gtrsim 1$, we estimate
$\varkappa \gtrsim 1$. In this case, the probability of the pair photoproduction is of the order of the
probability of emission of synchrotron photon by the electron.  Therefore, pair
production~(\ref{BreitWheeler}) should be also taken into account, that can be done
by means of Monte Carlo
technique~\cite{Nerush14} utilizing Baier--Katkov quasiclassical formulas~\citep{Baier98, Berestetskii82}.

Here we present the results
of numerical simulations of the Weibel instability in two counter-streaming hot and dense relativistic
plasma flows. Unlike synchrotron emission and pair production, particle collisions (e.g., Compton
scattering and bremsstrahlung) are not included in the simulations.

\begin{figure}
\includegraphics{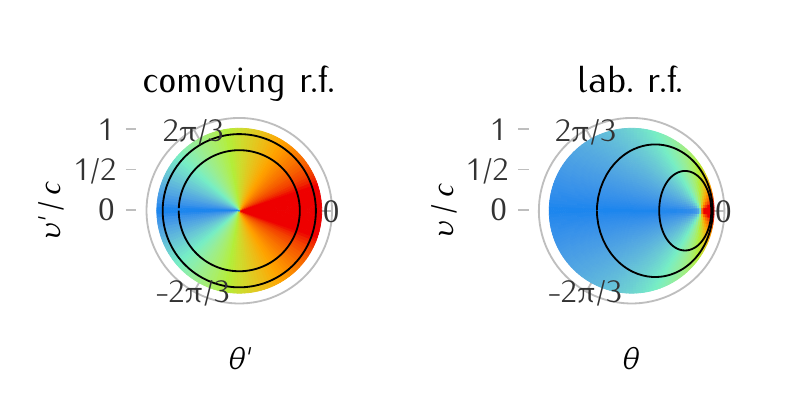}
    \caption{\label{theta} The distribution of the azimuth $\theta'$ in the proper reference frame
    of the flow ($K'$, left) and
    in the laboratory
    reference frame ($K$, right) (different values of $\theta'$ are shown with different colors;
    radial coordinate depicts velocity $v$).
    $\theta'=0$ corresponds to the direction of the relative velocity $V$ of the reference
    frames; for this plot $V = 0.86\,c$.}
\end{figure}

Let us also
note that in the theoretical considerations of the Weibel instability we follow electromagnetic
scenario~\citep{Stockem14}, because for ultrarelativistic flows
($\Gamma \gg 1$, where $\Gamma$ is the Lorentz factor of a
flow in some, e.g. in the laboratory, reference frame $K$), almost all velocity vectors of plasma particles belong to a
cone $\theta \lesssim 1 / \Gamma$ despite a high temperature of the flow (see Fig.~\ref{theta}; here $\theta$ is the
angle between the particle velocity and the flow velocity). This is true even if the average Lorentz
factor of the flow particles in
the comoving reference frame $K'$ is much greater than $\Gamma$, that is evident from the Lorentz transform of
angles from the proper reference frame of the flow $K'$ to $K$:
\begin{equation}
    \tan \theta = \frac{v_x'}{\Gamma (v_x' + V)} \tan \theta',
\end{equation}
where $V$ is the flow velocity in $K$ and the $x$ axis is parallel to it. Furthermore, it follows from the
transformation of the Lorentz factor:
\begin{equation}
    \label{gamma}
    \gamma = \gamma' \Gamma (1 + v_x' V),
\end{equation}
that the proper temperature of the flow determines the mean energy of particles in the
laboratory reference frame, $\bar \gamma = \bar \gamma' \Gamma$ (we assume $\overline{v_x'} = 0$).
Therefore, a hot plasma flow with
$\Gamma \gg 1$ should behave
similarly to a cold plasma flow, and the Weibel instability in the counter-streaming flows should grow in
accordance with the
electromagnetic scenario (formation and growth of current filaments with azimuthal magnetic field and
low electric field, see~\cite{Stockem14, Garasev17b} and references wherein) rather
than with the electrostatic one.

The paper is organized as follows. In Sec.~\ref{theory}, we consider the Weibel
instability of the electromagnetic type in counter-streaming relativistically hot plasma flows analytically, without synchrotron
emission and pair production. In Sec.~\ref{pairs}, we estimate the plasma parameters corresponding to
$\chi \sim 1$ and $\varkappa \sim 1$, hence, efficient synchrotron emission and copious pair production. In Sec.~\ref{num}, the
results of numerical simulations with synchrotron emission and pair production taken into account
are given, and in Sec.~\ref{conclusion}, their astrophysical implications are discussed.
In Sec.~\ref{summary}, the summary of the paper is given.

\section{Weibel instability in hot collisionless plasmas}
\subsection{Effect of temperature}
\label{theory}
Let us consider the stability of two relativistic counter-propagating plasma flows moving along the $x$ axis,
with respect to the formation of a cylindrically symmetric current filament.  The properties of the
flows are further denoted by indices $1$ (the flow velocity $v_x>0$) and $2$ ($v_x <0$). We also
assume that the filament remains quasineutral and
\begin{equation}
    \label{n}
    \delta n_{e 1} = \delta n_{i 2} = - \delta n_{i 1} = - \delta n_{e 2},
\end{equation}
where $\delta n$ is the density perturbation relative to the initial value for the flow
($n_1$ or $n_2$), the indices $i$ and $e$ refer to protons and electrons, respectively 
(the assumption~(\ref {n}) will be justified a bit later, it is not always fulfilled 
and is used to simplify calculations). Let $r$ and $\varphi$ be the cylindrical coordinates 
with respect to the axis of the current filament coinciding with the $x$ axis.
$x$, $r$ and $\varphi$ are right-handed coordinates. 
We also assume that none of the plasma characteristics depend on $x$, therefore 
the Maxwell's equations can be written as follows:
\begin{eqnarray}
    \label{dEdt}
    \frac{\partial E_x}{\partial t} = \frac{c}{r} \frac{\partial (r
    B_\varphi)}{\partial r} + 16 \pi e c \delta n_{e1}, \\
    \label{dBdt}
    \frac{\partial B_\varphi}{\partial t} = c \frac{\partial E_x}{\partial r}.
\end{eqnarray}

To obtain the equation for the density perturbation, 
one should start from the Boltzmann equation in Cartesian coordinates:
\begin{equation}
    \frac{\partial f}{\partial t} + (\mathbf v, \nabla) f +
    \frac{(\mathbf F, \nabla_{\mathbf v})}{m \gamma} f = 0
\end{equation}
where $f(r, \mathbf v)$ is the distribution function, $\gamma = (1 - v^2)^{1/2}$, and we assume
that for particles $\mathbf F \perp \mathbf v$ hence $d \gamma / dt = 0$.  Here the boldly denoted
vectors are in the $y-z$ plane, perpendicular to the $x$ axis. For particle density
\begin{equation}
     n(y,z) = \iiint_{v^2 < c^2} f(y, z, v_x, v_y, v_z) \, dv_x \, dv_y \, dv_z
\end{equation}
from the Boltzmann equation we obtain
\begin{equation}
    \label{dndt}
    \frac{\partial n}{\partial t} = -\nabla \cdot (n \bar {\mathbf v}),
\end{equation}
where the bar denotes averaging over velocities:
\begin{equation}
    \bar a = \frac{1}{n} \iiint_{v^2 < c^2} a f \, dv_x \, dv_y \, dv_z.
\end{equation}

For the average velocity we obtain
\begin{multline}
    \label{dvdt2}
    \frac{\partial \bar v_k }{\partial t} = \overline{\left( \frac{F_k}{m \gamma} \right)} - \bar v_l \frac{\partial
    \bar v_k}{\partial x_l} \\ - \frac{1}{n} \frac{\partial}{\partial x_l} \left[ \overline {( v_l
    - \bar v_l ) ( v_k - \bar v_k)} n \right],
\end{multline}
where Einstein summation convention is used.
We also assume that the covariance matrix for $\mathbf v$ is a constant, i.e.
\begin{equation}
    \mathcal V_{kl} \equiv \overline {(v_k - \bar v_k) (v_l - \bar v_l)} = \mathrm{const},
\end{equation}
and the distribution function is assumed to be symmetrical in the $yz$ coordinates, 
so that $\mathcal V_{k \neq l} = 0$ and $\mathcal V_{zz} = \mathcal V_{yy} \equiv \mathcal V$. Then
the equation~(\ref{dvdt2}) for $v_y$ or $v_z$ can be written as
\begin{equation}
    \label{dvdt3}
    \frac{\partial \bar v_k}{\partial t} = \overline{\left( \frac{F_k}{m \gamma} \right)} - \bar v_l \frac{\partial \bar
    v_k}{\partial x_l} - \frac{\mathcal V}{n} \frac{\partial n}{\partial x_k}.
\end{equation}

We suppose that in the case of a relativistically hot plasma in the proper reference frame, 
the plasma particles are uniformly distributed over the surface of a sphere
${v_x'}^2 + {v_y'}^2 + {v_z'}^2 \simeq c^2$, hence, using velocity transformation formulas
\begin{eqnarray}
    \label{vx}
    v_x &=& \frac{v_x' + V}{1 + v_x' V / c^2}\\
    \label{vy}
    v_{y, z} &=& \frac{v_{y, z}' \sqrt{1 - V^2 / c^2}}{1 + v_x' V / c^2},
\end{eqnarray}
one may easily derive the expression for $\mathcal V$ in the reference frame where the flow velocity is relativistic:
\begin{multline}
    \mathcal V \simeq \frac{1}{\Gamma^2} \iint_{{v_y'}^2 + {v_z'^2} < c^2} \frac{{v_y'}^2}{v_x' (1 + v_x' /
    c)^2} \, dv_y' \, dv_z' \\
    \times \left( \iint_{{v_y'}^2 + {v_z'^2} < c^2} \frac{1}{v_x'} \, dv_y' \, dv_z' \right)^{-1} \approx
    \frac{0.2 c^2}{\Gamma^2}.
    % b:=NIntegrate[1 / Sqrt[1 - x^2 - y^2], {x, -1, 1}, {y, -Sqrt[1 - x^2], Sqrt[1 - x^2]}]
    % t:=NIntegrate[x^2 / Sqrt[1 - x^2 - y^2] / (1 + Sqrt[1 - x^2 - y^2])^2, {x, -1, 1}, {y, -Sqrt[1 - x^2], Sqrt[1 - x^2]}]
    % t / b
\end{multline}
Here $v_x' = (c^2 - {v_y'}^2 - {v_z'}^2)^{1/2}$.

After that, the equations~(\ref{dndt}) and (\ref{dvdt3}) can be rewritten in cylindrical coordinates, 
assuming that $\bar v_\varphi = 0$:
\begin{eqnarray}
    \label{dndt2}
    \frac{\partial n}{\partial t} &=& -\frac{1}{r} \frac{\partial (r n \bar v_r)}{\partial r},\\
    \label{dvrdt}
    \frac{\partial \bar v_r}{\partial t} &=& \overline{\left(\frac{F_r}{m \gamma} \right)} - \bar v_r \frac{\partial \bar
    v_r}{\partial r} - \frac{\mathcal V}{n} \frac{\partial n}{\partial r}.
\end{eqnarray}

We consider only the initial stage of the instability, so for the force in Eq.~(\ref{dvrdt}) 
the following expression can be used:
\begin{equation}
    \bar F_r \simeq \pm e B_\varphi,
\end{equation}
where the sign is determined by the sign of $\bar v_x$ and the sign of the particle charge, hence we
can estimate
\begin{equation}
    \overline{\left(\frac{F_r}{m \gamma} \right)} \simeq \frac{F_r}{m \bar \gamma}.
\end{equation}
It can be noted that the sign of the force is the same for the ions (electrons) 
of the first flow and electrons (ions) of the second flow, so in the case of flows with the same
parameters (density, Lorentz factor and temperature) the density is perturbed such that the
quasineutrality condition~\eqref{n} stands true. Otherwise, when densities or Lorentz factors 
of the flows do not coincide, the condition~\eqref{n} may not be fulfilled, 
but we will use it for the sake of simplicity, assuming the plasma is quasi-neutral.

We look for the solution of the Maxwell's equations~(\ref{dEdt}),~(\ref{dBdt}) 
together with the equations~(\ref{dndt2}),~(\ref{dvrdt}) in the following form:
\begin{eqnarray}
    E_x &=& E_0 e^{\Lambda t} J_0(r / \lambda),\\
    \delta n_{e1} &=& -\delta n_0 e^{\Lambda t} J_0(r / \lambda),\\
    v_r &\propto& B_\varphi \propto e^{\Lambda t} \frac{d J_0(r / \lambda)}{dr},
\end{eqnarray}
where $E_0$ and $\delta n_0$ are the amplitudes, $J_0$ is the zero-order Bessel function 
of the first kind, i.e. the solution of the equation
\begin{equation}
    \frac{1}{r} \frac{d}{dr}\left\{ r \frac{d \left[ r J_0(r / \lambda) \right]}{dr} \right\} = -
    \frac{1}{\lambda} J_0(r / \lambda).
\end{equation}
Therefore, we obtain the equations which describe the parameters of cylindrically-symmetric modes:
\begin{equation}
    \label{main1}
    \left(1 - \frac{4 \omega_1^2}{\bar \gamma_1 \Lambda^2}\right) \frac{c^2}{\lambda^2} E_0 + \Lambda^2
    E_0 + \frac{16 \pi e c \mathcal V_1}{\Lambda \lambda^2} \delta n_0 = 0,
\end{equation}
\begin{equation}
    \label{main2}
    \frac{\mathcal V_1}{\lambda^2} \delta n_0 - \frac{n_1 e c}{\Lambda m \bar \gamma_1} \frac{1}{\lambda^2} E_0
    + \Lambda^2 \delta n_0 = 0,
\end{equation}
where $n_1$, again, is the first flow initial density $n_1 \equiv n_{e,1}(t = 0)$ and $\omega_1^2 = 4 \pi
e^2 n_1 / m$ is the related plasma frequency.

The first equation at $\mathcal V = 0$ describes, in addition to the stable mode, the Weibel instability,
and the second at $E_0 = 0$ describes quasi-sound waves. In the first and second cases, 
it is easy to obtain a relation between the characteristic spatial scale of the mode $\lambda$ 
and the characteristic "increment" $\Lambda$:
\begin{eqnarray}
    \label{E0}
    \Lambda^2_{\mathcal V = 0} & = & \frac{c^2}{2 \lambda^2} \left( -1 \pm \sqrt{
    1 + \frac{16 \lambda^2 \omega_1^2}{c^2 \bar \gamma_1}} \right),\\
    \label{n0}
    \Lambda^2_{E_0 = 0} & = & -\frac{\mathcal V_1}{\lambda^2}.
\end{eqnarray}

The relation between $\Lambda$ and $\lambda$ can be found from the 
equations~(\ref{E0}) and (\ref{n0}) in the general case as well:
\begin{equation}
    \Lambda^4 + \frac{c^2 + \mathcal V_1}{\lambda^2} \Lambda^2 - \frac{c^2}{\lambda^2} \left( \frac{4
    \omega_1}{\bar \gamma_1} - \frac{\mathcal V_1}{\lambda^2} \right) = 0,
\end{equation}
therefore, taking into account that $\mathcal V_1 \ll c^2$, we derive for the unstable mode
\begin{equation}
    \Lambda^2 = \frac{c^2}{2 \lambda^2} \left( -1 + \sqrt{1 + \frac{4 \lambda^2}{c^2} \left( \frac{4
    \omega_1^2}{\bar \gamma_1} - \frac{\mathcal V_1}{\lambda^2} \right)} \right).
\end{equation}
In the above equation it can be seen that at $\mathcal V \neq 0$ the considered mode is unstable
($\Lambda^2 > 0$) if
\begin{equation}
    \lambda > \frac{\sqrt{\bar \gamma_1 \mathcal V_1}}{2 \omega_1},
\end{equation}
i.e., for modes with a spatial scale, greater than some. It can be easily shown that in the presence
of temperature the maximum increment $\Lambda_m$ is realized for the mode with the following
spatial scale:
\begin{equation}
    \label{lambda_m}
    \lambda_m^2 \sim \mathcal V_1 \bar \gamma_1 / \omega_1^2, \quad \frac{\lambda_m}{\lambda_1} \sim
    \frac{1}{2 \pi} \frac{\bar \gamma_1^{1/2}}{\Gamma_1},
\end{equation}
and equals
\begin{equation}
    \label{Lambda_m}
    2\pi \Lambda_m / \omega_1 \sim 2 \pi / \sqrt{\bar \gamma_1}.
\end{equation}
Note that although we obtain Eqs.~\eqref{lambda_m} and \eqref{Lambda_m} for the flows with equal
parameters, we will use these equations for the flows with different parameters as well, assuming
that the index $1$ denotes the flow with higher corrected plasma frequency: $\omega_1 / \bar
\gamma_1^{1/2} > \omega_2 / \bar \gamma_2^{1/2}$ that yields higher value for the
increment~\eqref{Lambda_m}. The obtained
estimates are compared with results of numerical simulations in Sec.~\ref{num}.

\subsection{Pair production}
\label{pairs}

Here we consider pair photoproduction~(\ref{BreitWheeler}) in the electromagnetic fields
during the Weibel instability.
The pair production becomes efficient if $\varkappa \gtrsim 1$, where the quantum parameter $\varkappa$ depends 
on the field magnitude and the energy of the photon~(\ref{varkappa}). In order to
check if the process~\eqref{BreitWheeler} appears in some astrophysical objects, the magnitude of
electromagnetic fields and photon energy should be found.

%Из закона преобразования
%гамма-фактора~(\ref{gamma}) (он может быть получен из закона преобразования скоростей~(\ref{vx}),
%(\ref{vy})) можно получить, что плотность потока импульса
%для потока пропорциональна $\Gamma n V^2 \eta$.
For the sake of simplicity we consider the Weibel instability in two counter-streaming plasma flows in
the reference frame where the momentum flow is the same for both jets:
\begin{equation}
    \label{center_of_momentum}
    n_1 \Gamma_1 V_1^2 \eta_1 \simeq n_2 \Gamma_2 V_2^2 \eta_2.
\end{equation}
Here we estimate $\bar \gamma_1 \approx \Gamma_1 \eta_1$ and $\bar \gamma_2 \approx \Gamma_2
\eta_2$. The parameter $\eta$ defines the average kinetic energy of ions in the reference frame comoving with the
flow as follows:
\begin{equation}
    \eta = \overline{(\gamma_i - 1)}.
\end{equation}
From here on we assume that flow $1$ is denser than flow $2$ ($n_1 > n_2$), and
in flow $2$, ions and electrons are more energetic than in flow $1$ ($\bar \gamma_2 \gtrsim
\bar \gamma_1$).

We assume that a sizable part of the initial energy of the flows is transferred to the energy
of electromagnetic fields, and the magnitude of the fields can be estimated as follows:
\begin{equation}
    B^2 \sim 8 \pi n_2 m c^2 \bar \gamma_2,
\end{equation}
where we additionally suppose that the volume occupied by the plasma is not changed much while the
filaments grow.
An electron in strong enough fields emits
photons with energy about its own energy (namely if $\chi \gtrsim 1$, see Sec.~\ref{intro}). Therefore,
in $\varkappa$ (Eq.~\ref{varkappa}) we can estimate the photon energy as follows
\begin{equation}
    \varepsilon_\gamma \sim m c^2 \bar \gamma_2,
\end{equation}
that leads to
\begin{equation}
    \label{varkappa2}
    \varkappa \sim \bar \gamma_2^{3/2} \sqrt{8 \pi n_2 r_e \lambda_C^2},
\end{equation}
where $r_e = e^2 / (mc^2)$ is the classical electron radius and $\lambda_C = \hbar / mc$ is the
Compton wavelength. Supposing that the average electron energy initially or after the acceleration
process~\citep{Silva03, Spitkovsky08a} is as high as the initial ion energy, we have $\bar \gamma_2
\approx \Gamma_2 \eta_2 M / m$. Therefore, copious pair production is ensured if
\begin{equation}
    \label{criterion}
    \varkappa \sim (\eta_2 \Gamma_2 M / m)^{3/2} \sqrt{8 \pi n_2 r_e \lambda_C^2} \gtrsim 1.
\end{equation}
Here, again, all values are given in the center-of-momentum reference
frame~(\ref{center_of_momentum}) and the index $2$ denotes the flow whose particle density is lower
than the density of the other.

In the case of strong synchrotron losses the equipartition assumption can lead to
an overestimation of the fields magnitude.  On the other hand, we estimate the photon energy using the
mean particle energy and not taking into account high-energy spectrum tails~\citep{Silva03,
Spitkovsky08a}.
Thus, the resulting criterion of copious pair production Eq.~\eqref{criterion} remains relevant, as shown
in the next Section by means of numerical simulations.

\section{Results of numerical simulations}
\label{num}
To verify the above estimates, we performed three-dimensional numerical simulations of the development 
of the Weibel instability in counter-propagating hot plasma flows using the particle-in-cell (PIC)
code \textsc{quill}~\citep{Nerush2010VANT, Serebryakov15}.
The simulations were carried out taking into account emission of hard photons and pair
photoproduction
in a strong field using the Monte Carlo method~\citep{Elkina11,
Nerush14}.
Collisions of particles and, in particular, Compton scattering and bremsstrahlung are not taken into account. 
To solve the Maxwell equations and to approximate currents and fields, we used algorithms 
of~\cite{Pukhov99}, to solve the equations of motion we used the method of~\cite{Vay08}.

We chose the following simulation parameters: the size of the simulation region was $54 \times 24
\times 24 \, \lambda_1^3$, where $\lambda_1$, as before, is the plasma wavelength of the
denser flow.
Initially each of the flows occupied half of the region. 
The transverse step of the numerical grid was equal to $\Delta y = \Delta z = 0.14\,\lambda_1$, 
the longitudinal one was $\Delta x = 0.063\,\lambda_1$, the time step was $\Delta t = 0.06 \times 2 \pi /\omega_1$. 
The initial number of quasiparticles of each species (electrons and ions) in a cell was equal to $8$.
The quasiparticle merging algorithm~\cite{} was not used. The plasma density of the flows had a flat
transverse profile with a decrease in the density at the edges to zero on the scale $\sim 2\,\lambda_1$.
We used open boundary conditions that allowed the free outflow of the electromagnetic waves and
particles at the 
boundaries~\citep{Pukhov99}.

Initially, the particles of the flow in the comoving reference frame had the distribution:
\begin{eqnarray}
    f_i \propto e^{-(\gamma - 1) / \eta}, \\
    f_e \propto e^{-m (\gamma - 1) / (M \eta)},
\end{eqnarray}
hence the average kinetic energy of the ions (or the electrons) in the comoving reference frame was
equal to $M c^2 \eta$.

\begin{deluxetable*}{CCCCCCCCCCCCC}
    \tablecaption{Simulation parameters and results. \label{parameters}}
    \tablecolumns{13}
    \tablehead{
        \colhead{Simulation} &
        \colhead{$n_1$} &
        \colhead{$\eta_1$} &
        \colhead{$\Gamma_1$} &
        \colhead{$n_2 / n_1$} &
        \colhead{$\eta_2$} &
        \colhead{$\Gamma_2$} &
        \colhead{$M / m$} &
        \colhead{$\lambda_m / \lambda_1$} &
        \colhead{$2 \pi \Lambda_m / \omega_1$} &
        \colhead{$\varkappa$} &
        \colhead{$N_p / N_e$} &
        \colhead{$dN_p / dN_\gamma$} \\
        \colhead{} & \colhead{$(\text{cm}^{-3})$}
    }
\startdata
s1      & 1 \times 10^{25}   & 2   & 25 & 0.25 & 20  & 10  & 10 & 0.14 & 0.28 & 14   & 1.4 \times 10^{-3} & 6 \times 10^{-3} \\
s2_{26} & 6.3 \times 10^{23} & 20  & 10 & 1    & 20  & 10  & 10 & 0.71 & 0.14 & 7.2  & 1.4 \times 10^{-3} & 3 \times 10^{-3} \\
s3      & 2.5 \times 10^{24} & 5   & 10 & 0.4  & 5   & 25  & 1  & 0.11 & 0.89 & 0.14 & 1.5 \times 10^{-7} & < 10^{-6}        \\
s4_{22} & 1.6 \times 10^{23} & 1.3 & 10 & 1    & 1.3 & 10  & 15 & 0.22 & 0.45 & 0.11 & 1.4 \times 10^{-8} & <10^{-6}         \\
s5      & 7.7 \times 10^{22} & 10  & 16 & 0.5  & 2   & 160 & 15 & 0.49 & 0.13 & 6.5  & 7 \times 10^{-6}   & 7 \times 10^{-5} \\
s6_{23} & 1.9 \times 10^{22} & 7   & 4  & 0.7  & 4   & 10  & 20 & 0.94 & 0.27 & 0.27 & 3.6 \times 10^{-9} & <10^{-6}         \\
s7      & 5 \times 10^{24}   & 2   & 4  & 0.08 & 1.2 & 80  & 20 & 0.5  & 0.5  & 5.7  & 3.8 \times 10^{-6} & 1 \times 10^{-4} \\
\enddata
\end{deluxetable*}

We have carried out a series of seven simulations for different Lorentz factors, densities and
temperatures of the plasma flows.
The simulations parameters are given in Table~\ref{parameters}, where s$^*$ means the simulation
identifier. The proton to electron mass ratio $M / m$ in the simulations was chosen much lower than
that for the real particles in order to reduce computational costs.
For the given parameters of the simulations the instability growth rate $\Lambda_m$ and the
transverse scale of the filaments $\lambda_m$ were computed with Eqs.~\eqref{Lambda_m} and
\eqref{lambda_m}, respectively.
The parameter $\varkappa$ crucial for the pair photoproduction was estimated with Eq.~\eqref{varkappa2}.
In most simulations, the end time was equal to $t_{end} = 27 \times 2 \pi / \omega_1$.
However, in some simulations we were forced to terminate them before $t_{end}$ due to the significant growth of the number of particles (mostly photons).
In those simulations, the end time is given as a subscript in a simulation identifier (e.g., $s4_{22}$).
The ratio of the number of positrons to the number of electrons $N_p / N_e$
at the end of a simulation and the quantity $dN_p / dN_\gamma$ characterizing the
positron generation efficiency (see more details further) are computed in the simulations and are
also given in the Table~\ref{parameters}.

\begin{figure}
\includegraphics{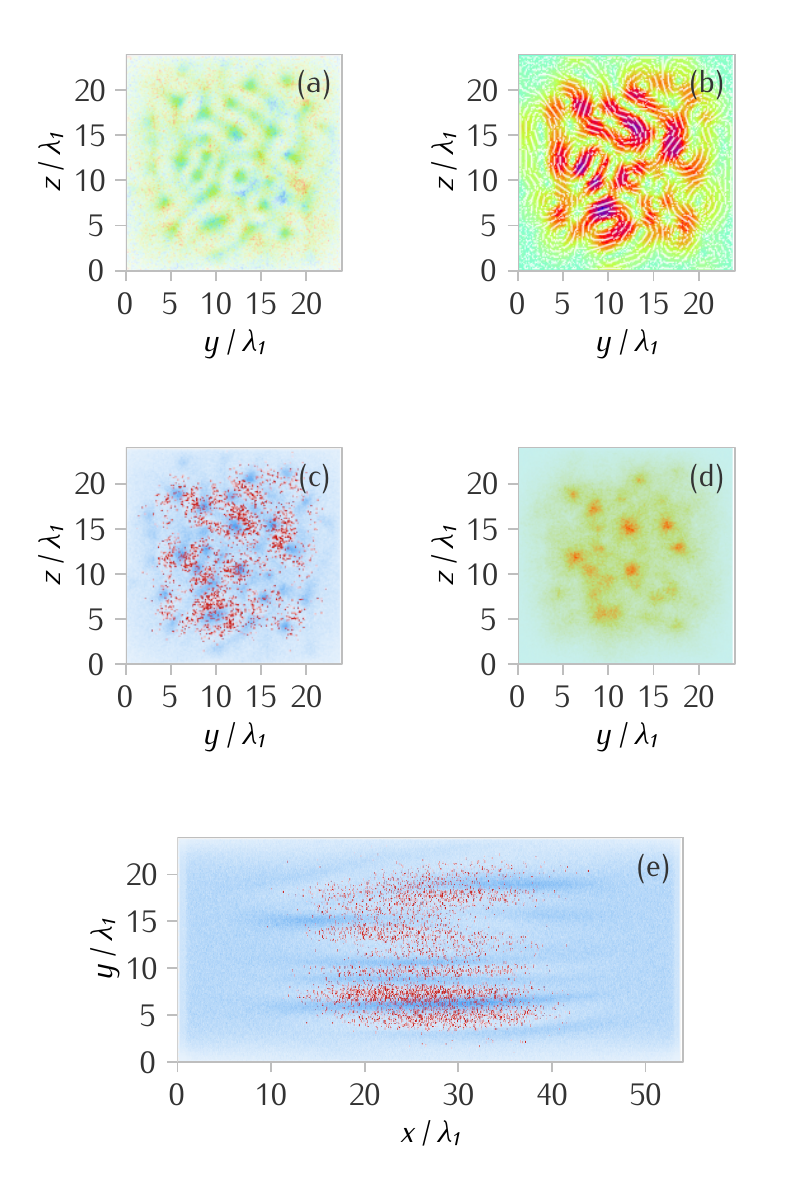}
\caption{\label{density}The results of s2 simulation. 
(a) In the $y-z$ plane, the total proton and electron density shown with the color intensity, the ratio of the charge
density to the total particle density depicted by the color hue (red color corresponds to a
plasma consisting of 60\% ions and 30\% electrons, blue color corresponds to 30\% ions and
60\% electrons mixture). 
(b) The transverse magnetic field energy density ($B_y^2 + B_z^2$) distribution in the $y-z$ plane, the darker colors correspond to the higher energy density, white lines sketchily show the field direction.
    The electron (blue) and positron (red) density distribution in (c) $yz$ and (e) the $x-y$ planes, respectively. 
Note that the maximum positron density is about 2 orders of magnitude lower than that of electrons.
(d) The gamma quanta density in the $y-z$ plane.
All distributions are given at the time instant $t = 26 \lambda_1 / c$.
Both the $x-y$ plane and the $y-z$ plane pass through the center of the simulation area.}
\end{figure}

Let us consider s2 simulation as an example. Fig.~\ref{density} (a) shows the sum electron
and ion density (as color intensity) as well as the relative electric charge (shown as color hue)
at the filaments cross-section. It is seen that plasma remains close to neutral during the
instability growth.
Fig.~\ref{density} (b) depicts the transverse (azimuthal) magnetic field generated around the filaments.
It should be noted that the typical filament size and the scale of the magnetic field they generate
is approximately of the order of the distance between the filaments.
From Figs.~\ref{density} (c) and (d) showing electron and photon density distributions, respectively, one can see
that the positions of these distributions maximums coincide. 
At the same time, the distribution of the generated positrons is
similar to the distribution of the magnetic field (see Figs.~\ref{density} (b), (c) and (e)).

\begin{figure}
\includegraphics{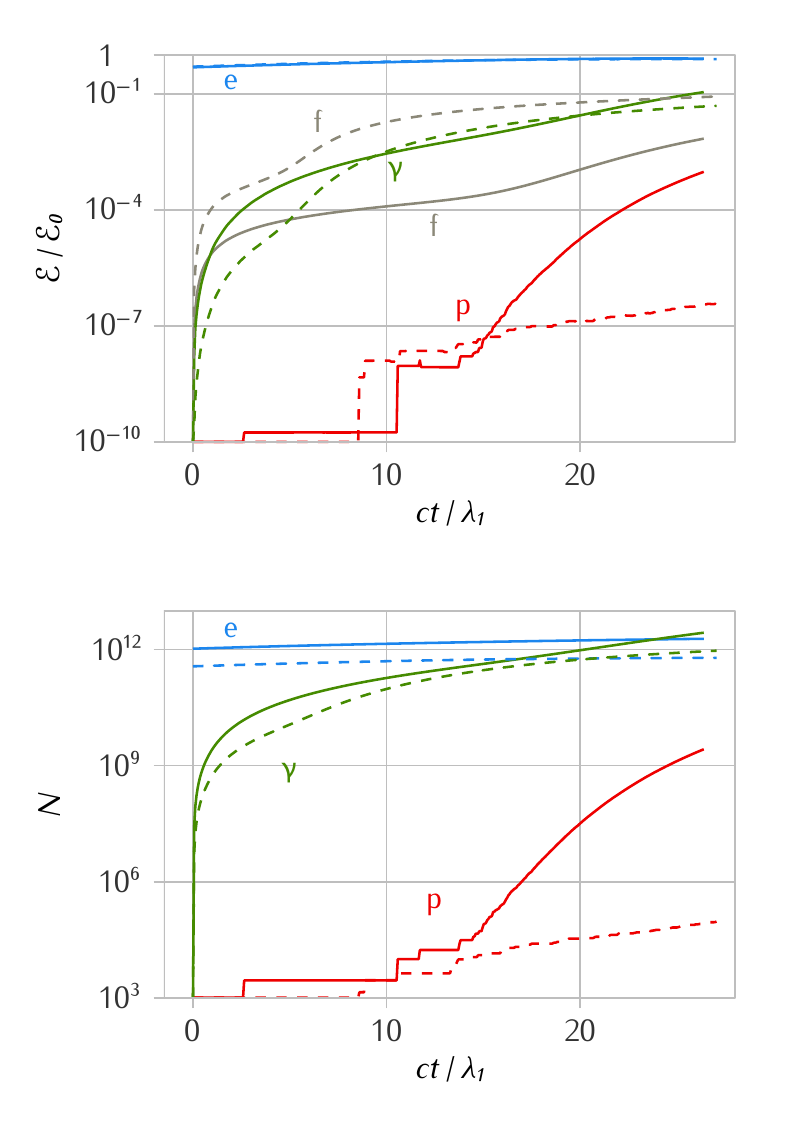}
\caption{\label{energy}The normalized particle and field energy (upper plot) and the particle
    number (lower plot) as functions of time in 2 different simulations: s2 (solid lines) and s3
    (dashed lines).  Electrons (e), positrons (p), photons ($\gamma$) and electromagnetic fields
    (f) are shown.}
\end{figure}

\begin{figure}
\includegraphics{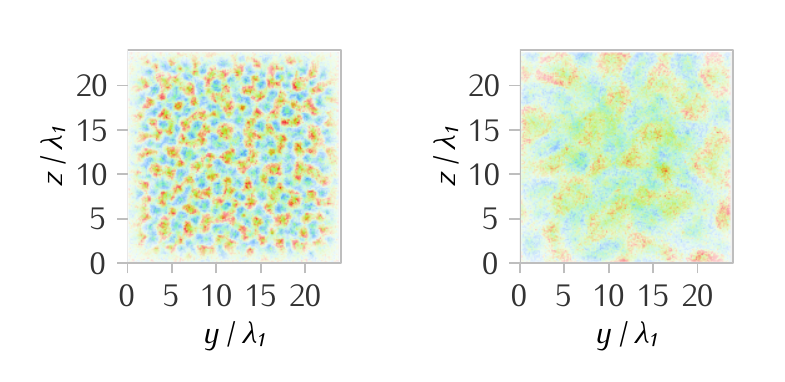}
\caption{\label{fmerging}The total electron and ion density with the relative charge density (as defined in Fig.~\ref{density}(a)) for s3 simulation at different time instants: $t = 10 \lambda_1 / c$ (left) and $t = 20 \lambda_1 / c$ (right).}
\end{figure}

In Fig.~\ref{energy} (top), the growth of the energy of electromagnetic fields, the energy of photons
and positrons in the process of instability development in s2 simulation is shown with solid
lines (for comparison, the dashed lines show the same quantities for s3 simulation). 
Figure~\ref{energy}(bottom) depicts the number of particles in the s2 and s3 simulations as a function of time. 
Despite the fact that the energy of electromagnetic fields in s3 simulation is higher, the number
of positrons produced in it is negligible compared to s2 simulation.  It can be seen from
Fig.~\ref{energy} (top) that the growth rate of the plasma fields energy (i.e.
the slope of the f lines)
depends on time, which is explained by the transition from the linear stage of development
of the instability to the nonlinear one. 
The nonlinear stage is characterized not only by the growth of fields and perturbation of the
plasma density, but also by the merging of current filaments. For an example, see the density
distribution in the $y-z$ plane for s3 simulation in Fig.~\ref{fmerging} at two different time instants.

\begin{figure*}
\includegraphics{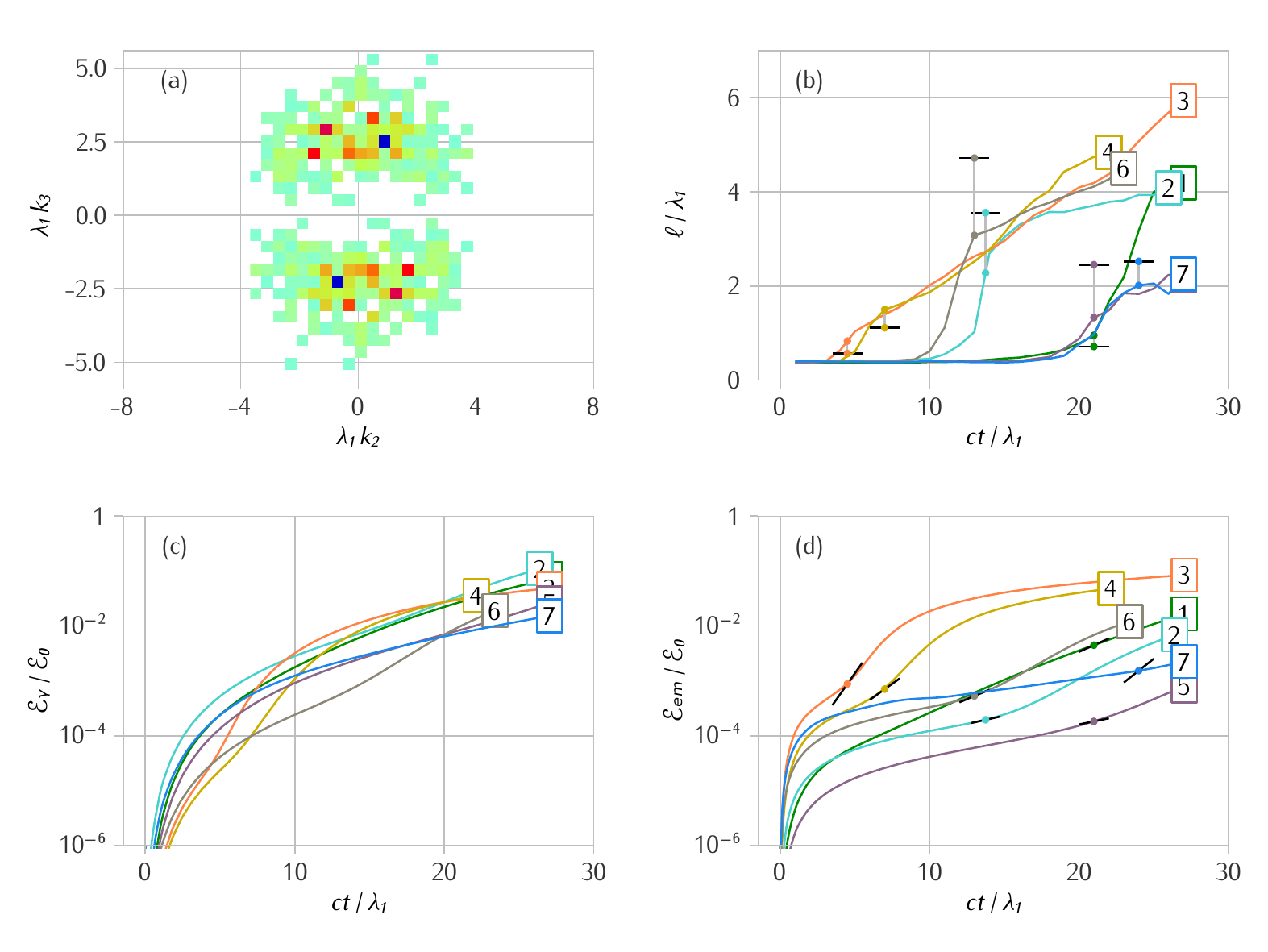}
\caption{\label{table-energy}(a) The two-dimensional Fourier image of the magnetic field component
    $B_y(y, z)$ in the simulation s3 at $t = 10 \lambda_1 / c$ and $x$ passing the center of the
    simulation box; $k_2$ and $k_3$ are the wavenumbers along the $y$ axis and the $z$ axis respectively. (b)
    Filament size $\ell$ for s1--s7 simulations determined from the Fourier images of $B_y$, as
    functions of time. Dots mark the time instances at which the evolution turns from linear to
    nonlinear stage. The ordinate of the black horizontal bars corresponds to the estimate of the
    filament scale~\eqref{lambda_m} multiplied by $5$, i.e. $5 \lambda_m / \lambda_1$ (see
    Table~\ref{parameters} for numerical values). (c) The energy of the gamma-rays and (d) the
    energy of the large-scale electromagnetic fields normalised to the initial ion energy of the
    plasma that will fill up the simulation box at the end of the simulation, $\mathcal E_0$. The
    short black lines are the exponents $\mathcal E_{em} \propto \exp{\Lambda_m t}$, where
    $\Lambda_m$ is the estimate of the instability growth rate~\eqref{Lambda_m} (see
    Table~\ref{parameters} for numerical values).}
\end{figure*}

Consider the entire set of the simulation results (s1--s7). The characteristic transverse scale of
the filaments $\ell$ was found from the simulation results as follows. First, the modulus of the Fourier image of
$B_y$ was computed, and the background of it (values below $0.1$ of its maximum) was deleted.
Fig.~\ref{table-energy} (a) shows such a Fourier image for s3 simulation and $t = 10 \lambda_1 /
c$. Then, using this image, the dispersion of the transverse wave vectors was computed, for
example, Fig.~\ref{table-energy} (a) yields the dispersion $k^2 \approx 9.7 / \lambda_1^2$ and
therefore $\ell = 2 \pi / k \approx 2 \lambda_1$. 

For s1--s7 simulations, the characteristic distance between filaments $\ell$ computed with this
method as function of time is given in Fig.~\ref{table-energy} (b). The dependence of the energy of hard
photons $\mathcal E_\gamma$ and the dependence of the energy of the electromagnetic field $\mathcal E_{em}$ on time are
depicted in Figs.~\ref{table-energy} (c) and (d), respectively. 

In small times, the magnetic field generated due to the Weibel instability is smaller than
the noise associated with the temperature, so the described method of filament scale computation
in small times gives a scale of the order of the transverse step of the numerical grid.
However, if the generated magnetic field becomes greater than the noise level, the sharp growth of
$\ell$ from these value to some other value occurs. We suppose that the value of $\ell$ computed at
the end of this sharp growth corresponds to filament scale reasonably well. The time instances of
this sharp growths and the resulting transverse scales of the filaments for s1--s7 simulations are shown in
Fig.~\ref{table-energy} (b) with dots, together with the estimated value of the filament size $5
\lambda_m$ computed with Eq.~\eqref{lambda_m} and marked with short black lines. We multiplied the analytical values $\lambda_m$ by 5
for better coincidence between theory and simulations. The need of this multiplier can be explained
by the fact that Eq.~\eqref{lambda_m} gives filament radius whereas the method of $\ell$ computation
gives the distance between filaments. Note that the filament size computed for s1 and s3
simulations is close to the step size of the numerical grid, thus, in these simulations the linear
stage of the Weibel instability was computed with higher inaccuracy than in the others.

The nonlinear stage of the development of the Weibel instability is characterized, first, by the
fact that the density perturbation becomes of the order of the initial particle density and,
second, the filament merging. 
In Fig.\ref{table-energy} (b), almost for all simulations the nonlinear stage starts right after the
marked time instances and manifest itself as the subsequent growth of $\ell$.
By the order of magnitude, the increment at the linear stage of the instability development,
obtained in numerical simulation, is in good agreement with the increment
estimating by Eq.~\eqref{Lambda_m} (see Fig.~\ref{table-energy} (d) and Table~\ref{parameters}),
that does not take into account many
factors.
For example, in the case of essentially different parameters of the flow 1 and the flow 2, the difference in the density of
protons and electrons in filaments can be of the order of the particle density itself (see
Fig.~\ref{fmerging}). 
In addition, the energy of emitted gamma quanta can significantly exceed the energy of the
generated electromagnetic fields even at the initial stage of the instability development (see
Fig.~\ref{energy}).

At the saturation of the Weibel instability, in the case of counter streaming plasma flows, the
filament current is determined only by the plasma density, and the maximal magnetic field is about $B
\sim n_e \ell$. Therefore, the filament size is strongly coupled with the energy of the magnetic field.
Therefore, the synchrotron emission should also lead to less filaments radius, because the radiation
losses reduce the energy of the magnetic field.

For s2--s6 simulations, a noticeable increase in the instability increment is observed during the
transition to the nonlinear stage, but after that the increment can decrease because of the
filaments growth
and the rise of the characteristic distance between them.
It should also be noted that a rapid change in the filament configuration at the nonlinear stage
(filament merging) can lead to the appearance of strong electric fields.

\begin{figure}
\includegraphics{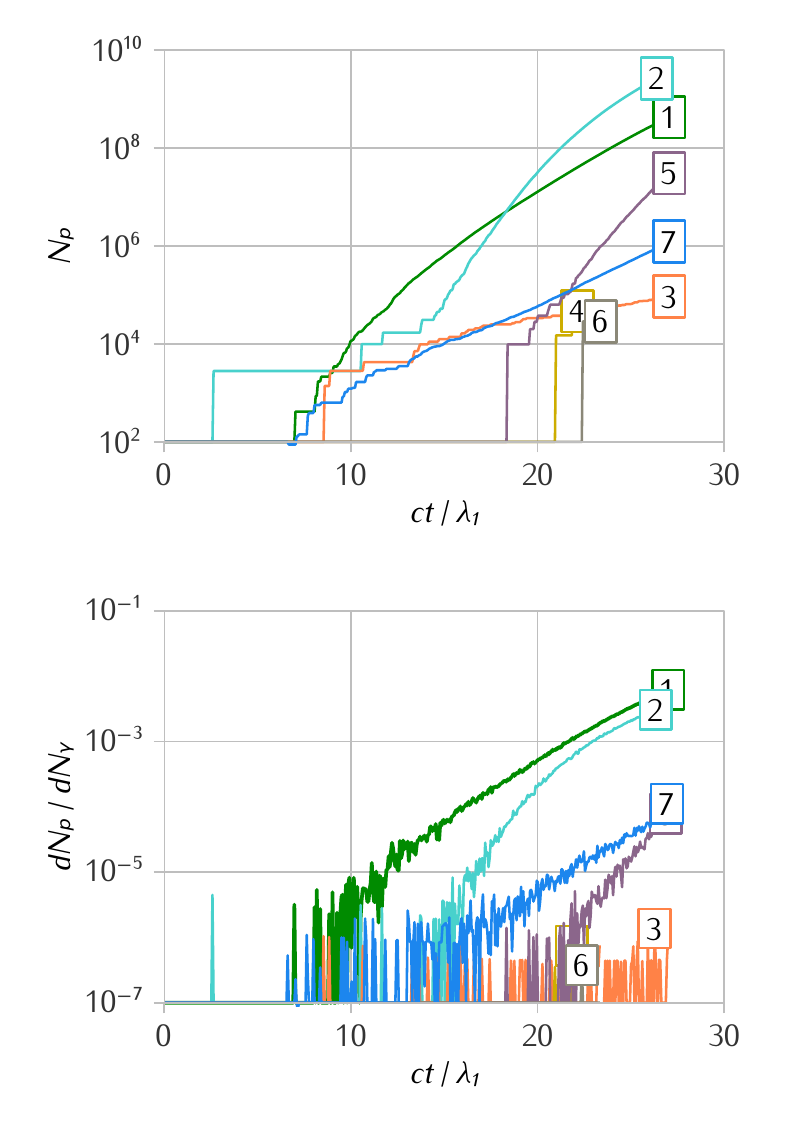}
\caption{\label{table-Np}(a) The number of positrons $N_p$ and (b) the parameter $dN_p / dN_\gamma$
    in s1--s7 simulations, as functions of time.}
\end{figure}

Numerical simulations in this work was carried out at the limit of technical capabilities available to the authors. 
Several calculations were stopped at $t<27 \lambda_1/c$ (until the flows intersected each other completely in the simulation region), because the of the large number of newly-born photons and limited RAM resources. 
Because of this, the saturation of the Weibel instability was not attained in almost all
calculations, however, in all calculations a nonlinear stage of instability was achieved (see
Fig.~\ref{table-energy}). 
Since the simulation parameters are different and the simulation time is sometimes less
than desired, we introduced the parameter $dN_p / dN_\gamma = (dN_p / dt) / (dN_\gamma / dt)$
to isolate the simulations with abundant positron production.
This parameter roughly shows the proportion of photons which produce electron-positron pairs. 
Dependences of the number of positrons and $dN_p / dN_\gamma$ on time in the s1--s7 simulations are
shown in Figs.~\ref{table-Np} (top) and (bottom), respectively.
From $N_p$ and $dN_p / dN_\gamma$ at the end of the simulations (see Table~\ref{parameters}), we
conclude that in the simulations s1, s2, s5 and s7, a significant production of
electron-positron pairs is realized. 
In the s3, s4 and s6 simulations, a low number of positrons is observed (despite the significant
number of the photons), and $dN_p / dN_\gamma$
does not exceed the background noise values. Thus, the criterion
(\ref{criterion}), yielding $\varkappa > 1$ for s1, s2, s5 and s7 simulations and $\varkappa < 1$
for s3, s4 and s6 simulations, does indeed allow us to distinguish the pair production regime during the
development of the Weibel instability.

\section{Discussion and astrophysical implications}
\label{conclusion}

In this paper, we consider the Weibel instability in two relativistic plasma flows that can lead to the
efficient synchrotron emission. The numerical simulations demonstrate that the conversion
efficiency of initial flows energy to the energy of synchrotron photons is much higher than
that for the generation of large-scale magnetic fields, if the flows are quite dense and energetic.

Numerical simulations also show that the synchrotron photons can produce $e^+e^-$ pairs in the
magnetic field,
giving the number of positrons up to $10^{-3}$ and higher of the number of
electrons in the flows. In order to clarify the flows parameters leading to copious pair
production, the theoretical estimate~\eqref{criterion} can be rewritten
using the rest-mass density of the hydrogen plasma of the flows $\rho$ (namely the density of the
cooled plasma in the comoving reference frame):
\begin{equation}
    %\varkappa \sim [\eta_2 \Gamma_2 M / m]^{3/2} \sqrt{8 \pi n_2 r_e \lambda_C^2},
    % H, 1 g / cm^3 -> 6.02 * 10^23 cm^{-3}
    % 1836**(3/2) * (sqrt $ 8 * pi * 6.02e23 * 2.82e-13 * 3.86e-11 ) = 6.2
    \label{criterion_conclusion}
    \varkappa \sim 6.2 \times \eta_2^{3/2} \Gamma_2^2 \sqrt{\rho_2[\text{g\, cm}^{-3}]} \gtrsim 1,
\end{equation}
where the flow Lorentz factors $\Gamma_{1,2}$ are given in the
center-of-momentum reference frame~(\ref{center_of_momentum}), $\eta = \overline{(\gamma' - 1)}$
is the mean normalized kinetic energy of the ions in the reference frame comoving with the flow. The
index $2$ denotes the flow whose particle density is lower than the density of the other, i.e. $n_2
\leq n_1$. For instance, this estimate yields $\varkappa \approx 1$ for $\eta_2 \sim 1$, $\Gamma_2 =
5$ and $\rho_2 \sim 10^{-4} \text{ g}\; \text{cm}^{-3}$.

The simulation results s1--s7 are obtained for $M / m$ far from the real proton-to-electron mass
ratio ($\approx 1836$), but can be scaled in a way that conserves the base estimate~(\ref{criterion})
as follows: $\eta_2 \Gamma_2$ from Table~\ref{parameters} is multiplied by $a M / (1836 m)$,
and $n_2$ is replaced by $n_2 / a^3$, where $a$ is an arbitrary constant (we choose $a = 10^{4/3}$
in order to fit $\eta_2 \Gamma_2$ in the range $1$--$100$). Values of $\eta_2 \Gamma_2$ and $n_2$ 
obtained with this scaling correspond to a hydrogen plasma and can be tested with
criterion~(\ref{criterion_conclusion}) and compared with believed values of these parameters for
astrophysical jets.

\begin{figure}
\includegraphics{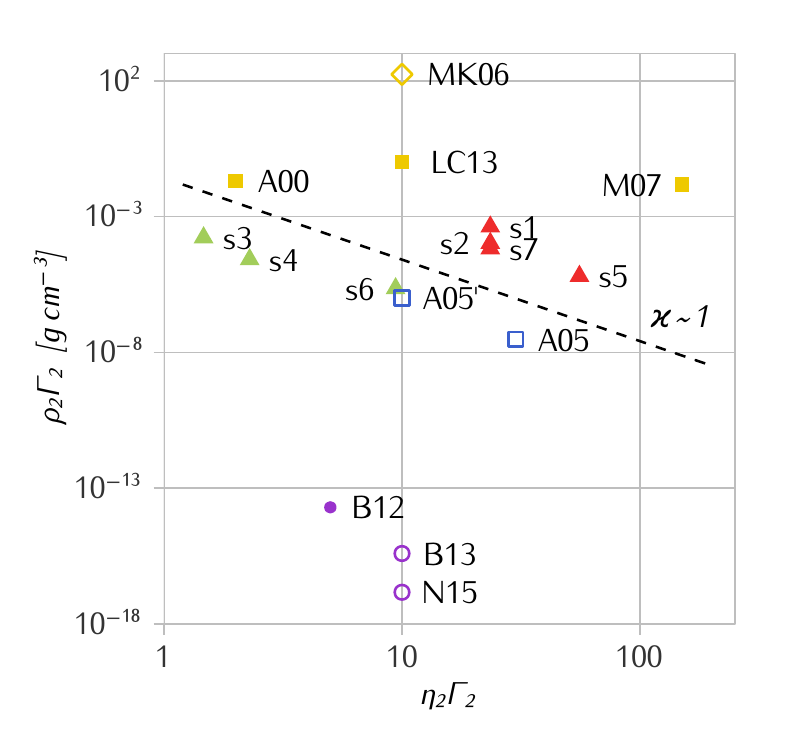}
  \caption{\label{gammaN} Parameters of counter-streaming plasma flows.
    A subset of simulations (this paper, red triangles s1, s2, s5, s7) demonstrates copious pair production and significant
    synchrotron losses, whereas in the other simulations of this paper (green triangles s3, s4, s6)
    the positron yield is low and the energy of synchrotron photons does not much exceed the energy of
    magnetic field. 
    These subsets evidently belong to the regions $\varkappa > 1$ (above the dashed
    line) and $\varkappa < 1$ (below it), respectively (for $\varkappa$ see
    Eq.~~\eqref{criterion_conclusion} and Table~\ref{parameters}). 
    A number of numerical models of
    GRBs (collapsars with neutrino-antineutrino annihilation powered jets (yellow squares A00, LC13, M07), mergers (hollow
    blue squares A05 and A05$'$), a collapsar with Blandford--Znajek powered jet (a hollow yellow diamond MK06) found in the literature, as well as the
    estimated properties of a tidal disruption event leading to jet formation (solid circle B12) and
    blazars (hollow circles B13 and N15) are also shown (see text for details).}
\end{figure}

The line corresponding to Eq.~\eqref{criterion_conclusion} and $\varkappa \sim 1$ along with points
obtained from the simulation results s1--s7, are shown in Fig.~\ref{gammaN}. Simulations s1, s2, s7
and s5, resulting in high number of positrons and high rate of their production, are marked with
red triangles. Simulations s3, s4 and s6, resulting in low number of positrons generated, are
marked with green triangles. It is clearly seen that the line $\varkappa \sim 1$ divides
well the regions of copious and weak positron production, and Eq.~\eqref{criterion_conclusion} can
be used to test various astrophysical objects.

\subsection{Gamma-ray bursts}

Dense relativistic plasma jets are often associated with gamma-ray bursts (GRBs), tidal disruption
events, active galactic nucleus and blazars. The energy of particles in the jets and the jet mass
density could not be measured directly, however, the values used in a number of models of this
phenomena can be used.

In the collapsar model of
MacFadyen \& Woosley~\citep{MacFadyen99, Woosley99}, GRBs are linked with rotating massive stars whose
core collapse produces black hole swallowing surrounding matter. 
In that process, strong jets are generated due to energy deposition in the progenitor star 
envelope within the cone region around the rotation axis of the star. 
This energy deposition can be associated with
neutrino-antineutrino annihilation with subsequent heating and acceleration of the baryonic matter. 
The Weibel instability can
rise either in internal shocks in the jets or external shocks with pre-explosive stellar wind or
the star envelope. Note that in this model huge external pressure that accelerates the jets is
often associated not with the ion temperature but mostly with radiation, hence we use $\eta
\sim 1$ for this model. 
Note that such assumption neglects $e^+e^-$ pairs produced by neutrino-antineutrino
annihilation and contributing to the plasma density, hence the parameter $\varkappa$ given
for the collapsar models below is rather underestimated.

In the simulations based on the MacFadyen \& Woosley model~\citep{Aloy00} with the energy deposition of
the order of $10^{50} \text{--} 10^{51} \text{ erg}$, the jet breaking-out the progenitor star has
the rest-mass density about $10^{-1} \; \text{g} \, \text{cm}^{-3}$, the temperature $\eta \sim 1$ 
and the Lorentz factor
$\Gamma \sim 5$, while the envelope of the star is motionless and has density about $1 \; \text{g}\,
\text{cm}^{-3}$ (see dotted lines in Fig.~2 from~\cite{Aloy00}). In the center-of-momentum reference frame,
the Lorentz factor of the less dense flow (the jet) can be estimated as follows:
$\Gamma_2 \sim 2$, while the rest-mass density and thermal energy of ions, obviously, are the same,
$\rho_2 \sim 10^{-2} \; \text{g} \, \text{cm}^{-3}$, $\eta_2 \sim 1$. These parameters yield $\varkappa \approx
2.5$ and are shown as yellow square A00 in Fig.~\ref{gammaN}.

In the two-dimensional simulation of~\citet{Morsony07} adhering the MacFadyen
\& Woosley collapsar model and a power law stellar envelope model, the energetic jet ($\Gamma
\approx 300$, $\rho \sim 10^{-4} \text{g} \, \text{cm}^{-3}$, see color version of Fig.~3
from~\cite{Morsony07}) breaks out the star envelope ($\rho \sim 10^{-1} \; \text{g} \, \text{cm}^{-3}$) that
provides favorable conditions for the extreme Weibel instability ($\Gamma_2 \sim 150$, $\eta_2 \sim
1$, $\rho_2 \sim 10^{-4} \; \text{g} \, \text{cm}^{-3}$, $\varkappa \approx 10^3$, see the red square M07 in Fig.~\ref{gammaN}). In the further
development of this model (three-dimensional simulation with more realistic stellar progenitor,
see~\cite{LopezCamara13}) the parameters of jet breaking out the progenitor star (at $t =
4.2 \; \text{s}$) are slightly different: the jet has $\Gamma \approx 10$ and $\rho \sim 10^{-2} \;
\text{g} \,
\text{cm}^{-3}$, and the envelope has $\rho \sim 1 \; \text{g} \, \text{cm}^{-3}$ (see the green lines in Figs.~4 and 6
of~\cite{LopezCamara13}). These parameters yield $\Gamma_2 \sim 10$ and $\rho_2 \sim 10^{-2}
\text{g}
\, \text{cm}^{-3}$ (shown in Fig.~\ref{gammaN} as LC13 yellow square) which, together with $\eta \sim 1$, is
above the threshold $\varkappa \sim 1$ ($\varkappa \approx 60$).

Short GRBs are not linked with supernova explosions, and it is proposed that mergers (neutron star ---
neutron star or neutron star --- black hole mergers) could be the source of such bursts. It implies lower
density of the ambient and the jet plasma, and greater Lorentz factor of jets in
general~\citep{Aloy05}. For instance, in the simulation B01 at time $0.5 \; \text{s}$ (see Figs.~25 and 26
of~\citet{Aloy05}), the Lorentz-factor of the jet head is $\Gamma \approx 1000$, and its rest-mass
density is only $\rho \sim 10^{-9} \; \text{g} \, \text{cm}^{-3}$. 
Assuming the internal shock in such jet having
$\Gamma_2 \sim {\Gamma}^{1/2} \approx 30$ (A05 hollow blue square in Fig.~\ref{gammaN}), we obtain
$\varkappa \approx 0.2$. Earlier, i.e. at time $0.1 \text{ s}$, the head of the jet
has the Lorentz-factor $\Gamma \sim 100$ and density $\rho \sim 10^{-7} \; \text{g} \, \text{cm}^{-3}$ (see Figs.~15
and 16 in~\cite{Aloy05}). The corresponding parameters of internal shock with $\Gamma_2 \sim
\Gamma^{1/2} \approx 10$ again are not favorable for pair production during the Weibel instability
($\varkappa \approx 0.2$) and are shown as A05$'$ hollow blue square in
Fig.~\ref{gammaN}.

Another model of jet formation in long GRB engines connects it with the Blandford--Znajek
mechanism of energy extraction from rotating black hole~\citep{Blandford77}, and predicts the formation of a magnetically
driven outflow~\cite{McKinney06} (i.e, an outflow with magnetic pressure dominating over particle
pressure and Poynting flux dominating over the flux of the particle energy). This model allows one to
estimate the plasma density if the GRB luminosity and the mass of the central black hole is known 
(see the next subsection for details). For example, for a black hole with mass $M_{BH} = 10 M_\odot$ 
(where $M_\odot$ is the solar mass) and the overall jet luminosity $L_j =10^{50} \; \text{erg} \,
\text{s}^{-1}$ that is
typical for long GRBs~\citep{Piran99} one can obtain huge density $\rho_2
\sim 17 \; \text{g} \, \text{cm}^{-3}$ that together with $\Gamma_2 \approx 10$ leads to $\varkappa
\sim 10^3$
and is shown with MK06 hollow yellow diamond in Fig.~\ref{gammaN}.

Thus, the pair production regime of the Weibel instability potentially can be reached in long
gamma-ray bursts associating with collapse of massive stars. Short gamma-ray bursts associating with
merging of black holes or neutron stars presumably provides $\varkappa \lesssim 1$ and negligible
rate of pair production in the magnetic field of collisionless shocks.

\subsection{Supermassive black holes}

It is generally believed that super massive black holes (SMBHs) drive energetic outflows in active
galactic nucleus (AGNs) and blazars. However, a large value of Schwarzschild
radius of SMBHs implies a low value of the plasma density and $\varkappa \ll 1$.

Swift~J164449.3+573451 source, which is associated with a tidal
disruption of a star by a dormant SMBH~\citep{Zauderer11}, is of a certain interest because the observable
data allows one to estimate the jet parameters not far from, but quite near the black hole. Rapid time
variability of the gamma-rays and X-rays requires a compact source with a characteristic size of 
$\lesssim 0.15 \; \text{AU}$
($\lesssim 2 \times 10^{12} \; \text{cm}$)~\citep{Berger12}. More than 200 days of radio observations of the source
let one obtain the jet properties at the distance $r_{rf} \sim 10^{18} \; \text{cm}$ from the black
hole~\citep{Berger12}: $\Gamma \approx 5$ and $n(r_{rf}) \sim {1} \; \text{cm}^{-3}$. Assuming that the opening
angle of the jet $\theta_j \sim 5^\circ$, the distance between the SMBH and the gamma- and X-ray
source is $r_\gamma \sim (\tan \theta_j)^{-1} \times 0.15 \; \text{AU} \sim 2 \times 10^{13} \;
\text{cm}$ that yields
at this distance $n(r_\gamma) \sim n(r_{rf}) r_{rf}^2 / r_\gamma^2 \sim 2.5 \times 10^9
\text{cm}^{-3}$, hence
$\rho_2 \sim 4 \times 10^{-15} \text{g} \, \text{cm}^{-3}$. This value, together with $\Gamma_2 \approx 5$ and $\eta_2
\approx 1$, gives $\varkappa \sim 10^{-5}$.
and is depicted as B12 violet circle in Fig.~\ref{gammaN}.

The parameters of blazar jets can be similarly found from radio observations and luminosity in all
bands, and then can be continued up to the distance closer to the central black hole. The distance
from the black hole $r_\gamma$, at which the internal shock and the Weibel instability rise, is crucial
for a plasma density estimate and can be found as follows. First, $r_\gamma$ is connected with
the variability timescale $t_{var}$ and the jet opening angle $\theta_j$, $r_\gamma \sim c t_{var}
(\tan \theta_j)^{-1}$. Second, the numerical hydrodynamical model of jet formation of~\citet{McKinney06} which takes into
account general relativity and is capable to model Blandford--Znajek mechanism of jet
supply~\citep{Blandford77}, predicts that the magnetic pressure dominates in the jet from the region
of jet formation up to the Alfven surface at $r_A \sim 10$--$100 \, r_g$, where $r_g = 2GM_{BH} / c^2$
is the black hole Schwarzschild radius, $M_{BH}$ is the black hole mass and $G$ is the
gravitational constant. Beneath the Alfven surface, the internal shocks are absent in the simulations
of~\cite{McKinney06}, hence $r_\gamma \geq r_A \sim 100 r_g$.

Let us assume that jet luminosity $L_j$ is equal to the jet energy traveling through the jet
cross-section at $r_\gamma$, and the particle energy becomes comparable with the energy of magnetic
field here, hence
\begin{equation}
    \label{Lj}
    L_j \sim \pi {r_\gamma}^2 \rho c^3 \eta \Gamma^2 \tan^2 \theta_j,
\end{equation}
Thus, in order to estimate $\rho$ one should know $M_{BH}$,
$\Gamma$ and $\eta$. Relying on the simulations of~\cite{McKinney06}, we use $\Gamma \approx 10$ and
$\theta_j \approx 5^\circ$ in
the further estimations, additionally assuming $\eta \sim 1$.

In order to estimate parameters of internal shock nearest to the black hole of the famous blazar
3C273, we follow~\cite{Bottcher13, Zdziarski15}. In the leptonic model of~\cite{Bottcher13}. $L_j
\approx 1.3 \times 10^{46} \; \text{erg} \, \text{s}^{-1}$ (see Eq.~5  and value of $L_p$ in Table~2 wherein), and in Ref.~\cite{Zdziarski15}
the black hole mass is assumed to be $M_{BH} \approx 7 \times 10^9 M_\odot$, that yields $r_\gamma
= 2 \times 10^{15} \; \text{cm}$, $\rho \approx 4 \times 10^{-17} \; \text{g} \, \text{cm}^{-3}$ and $\varkappa \sim 10^{-6}$ (see B13
hollow blue circle in Fig.~\ref{gammaN}). Note that the variability
timescale $t_{var} \sim 1 \; \text{day}$ gives a slightly higher value of $r_\gamma \sim 3 \times
10^{16} \; \text{cm}$ and even lower
value of $\varkappa$.

The reported detection of gravitational lensing of the blazar PKS 1830-211~\citep{Neronov15}
independently provides the size of the gamma-ray emitting region about $r_\gamma / \tan \theta_j
\sim 10^{15} \text{cm}$, that coincides
fairly well with about $1 \; \text{day}$ variability timescale and $10$--$100 \; r_g$ for the central black
hole~\citep{Neronov15}. Thus, we adopt $r_\gamma \sim 10^{16} \; \text{cm}$ that, together with luminosity
$L_j \sim 3 \times 10^{45} \; \text{erg} \, \text{s}^{-1}$, leads to $\rho \sim 1.5 \times 10^{-18}
\; \text{g} \, \text{cm}^{-3}$ and $\varkappa
\sim 10^{-6}$ (see N15 violet hollow circle in Fig.~\ref{gammaN}).

Therefore, SMBHs provides
outflows with very low plasma density and $\varkappa \ll 1$.

\subsection{Collisions}

Fig.~\ref{gammaN} clearly demonstrates that copious emission of hard photons and pair production
during Weibel instability rises if the plasma density is at least $10^{-8} \; \text{g} \,
\text{cm}^{-3}$. In such plasmas, electron-photon and electron-ion collisions can be important, and
the corresponding cross-sections should be estimated.

Compton scattering cross-section in the center-of-momentum reference frame can be estimated as
follows~\citep{Berestetskii82}:
% ~section 86, p. 404
\begin{equation}
    \sigma_C \sim \frac{r_e^2}{\bar \gamma^2} \ln {\bar \gamma},
\end{equation}
where $r_e = e^2 / (mc^2)$ is the classical electron radius, the electron and photon energies are
approximately equal to each other and to ${\bar \gamma
} m c^2$. Thus, the ratio of the free time $t_f^{(C)} = 1 / n c \sigma$ (the mean time between two scattering events of the
same particle) to the Weibel instability timescale $\Lambda_m^{-1}$~(\ref{Lambda_m}) is:
\begin{equation}
    \Lambda_m t_f^{(C)} \sim \frac{\bar \gamma^{3/2}}{\ln \bar \gamma}
    \frac{\lambda}{r_e} \gg 1
\end{equation}
for almost any realistic plasma density (here $\lambda$ is the plasma wavelength).

Electron-proton scattering can be considered similarly. The momentum-transfer (transport) cross
section $\sigma_{mt}$ is
determined mostly by events with little change in the particle directions, and formulas for
electron scattering in a constant field can be used~\citep{LandauII, Berestetskii82, LandauI}:
% error in (80.10) - eps^2 should be used instead of eps^4.
% ~ section 80, p. 361, section 139, p. 697; Landau I, section 19.
% transport cross-section - see LL I, sec. 19.
% relativistic electron motion - LL II, sec. 39.
\begin{equation}
    \sigma_{mt} \approx \frac{8 \pi r_e^2}{\gamma_0^2} \ln
    \frac{\theta_{max}}{\theta_{min}},
\end{equation}
where $\theta_{max}$ and $\theta_{min}$ are the maximum and minimum deflection angles of the
electron trajectory, respectively, and $\gamma_0$ is the initial Lorentz factor of the scattered
electron. In the limit $\theta \ll 1$, the angles can be estimated as~\citep{LandauII}
\begin{equation}
    \theta \approx \frac{2 r_e}{\gamma_0 r_0 },
\end{equation}
where $r_0$ is the impact parameter. The minimal deflection angle $\theta_{min}$ can be estimated
using the Debye length $r_D \sim c \bar \gamma^{1/2} / \omega_{p}$, and, as far as the electron de
Broglie wavelength is smaller than the proton size ($\sim r_e$) and if $\gamma_0 \gtrsim \hbar c / e^2
\approx 137$, the maximal deflection angle can be estimated using the proton size.
Therefore, we estimate the momentum-transfer cross-section as follows:
\begin{equation}
    \sigma_{mt} \sim \frac{8 \pi r_e^2}{\gamma_0^2} \ln
    \frac{\lambda \bar \gamma^{1/2}}{r_e},
\end{equation}
and the ratio of the corresponding timescale $t_f^{(ei)}$ to the timescale of the Weibel
instability~\eqref{Lambda_m} as follows:
\begin{equation}
    \Lambda_m t_{f}^{(ei)} \sim \frac{\lambda}{r_e} \bar \gamma^{3/2} \ln^{-1}
    \frac{\bar \gamma^{1/2} \lambda}{r_e},
\end{equation}
This ratio is smaller than $\Lambda_m t_f$ by a logarithmic factor of the order of $10$, thus
$\Lambda_m t_f^{(ei)}$ is also much greater than unity for almost all plasma parameters.

The characteristic timescale of electron energy losses caused by bremsstrahlung is about the
timescale of $e^+e^-$ pair production by a photon colliding with a proton~\citep{Berestetskii82},
and is the following:
\begin{equation}
    \Lambda_m t_f^{(b)} \sim \frac{1}{\alpha \bar \gamma^{1/2} \ln 2 \bar \gamma} \frac{\lambda}{r_e},
\end{equation}
where $\bar \gamma$ is the Lorentz factor of the emitting electron or the energy of the
photon producing $e^+e^-$ pair, normalized to $m c^2$. For the density $\rho_\sim 10^{-4} \; \text{g} \,
\text{cm}^{-3}$ and $\bar \gamma = 1.8 \times 10^4$ providing $\varkappa \sim 1$, we
have $\Lambda_m t_f^{(b)} \sim 10^8$.

Therefore, for the parameters of interest the effect of collisions are negligible on the Weibel
instability timescale. However, at least the scale $c t_{f, b}$ is less than the size of a
gamma-ray emitting region in the collapsar model of GRBs ($r_\gamma$ is less or about $1$ light
second for a $\nu \tilde \nu$-annihilation driven jet and $r_\gamma \sim 10^7 \; \text{cm}$ for a
jet driven by Blandford--Znajek mechanism). Namely, for a photon density of the order of $n_1
\sim n_2$, and $\rho_{1,2} \sim 10^{-4} \; \text{g} \, \text{cm}^{-3}$, $\eta_{1,2} \sim 1$,
$\Gamma_{1,2} \sim 10$ we obtain $ct_f^{(C)} \sim 10^{13} \; \text{cm} \gg r_g$, $ct_f^{(ei)} \sim 10^{12} \;
\text{cm} \gg r_g$ and $ct_f^{(b)} \sim 10^{6} \; \text{cm} \ll r_g$.

Thereby the spectral energy distribution (SED) of photons would be drastically modified as they
disappear in the $e^+e^-$ photoproduction
in collisions with nucleus. The cross-section of this process for high-energy photons
($\bar \gamma \gg 1$) depend logarithmically on the photon energy, and the threshold of
the pair photoproduction $\bar \gamma \sim 1$ should be distinguished in the SED. Indeed, Fermi
GBM data demonstrate that most SEDs of the detected GRBs have a break in the power-law
fit~\citep{Gruber14} or maxima in the photon energy distribution~\cite{Abdo09} at $100$--$1000 \;
\text{keV}$. The maximal photon energy detected in GRBs (tens of GeV, see~\citet{Ackermann14,
Abdo09}) is about the energy of a proton with Lorentz factor about $100$ that coincides well with
the generally believed Lorentz factor of GRB jets. Anyway, the generation of observed high-energy
photons hardly can be attributed to high-density shock-wave region because of complicated
energy-temporal distribution of photons~\citep{Ackermann14, Abdo09}. Moreover, blazars also emitting
photons with energy $\sim 10 \; \text{GeV}$ nevertheless they have no regions of high-density plasma
(see Fig.~\ref{gammaN}) that implies other mechanisms of high-energy photons generation (e.g.,
comptonisation).

Thus, collisional effects are negligible on the timescale of Weibel instability, however,
bremsstrahlung as well as pair production in photon-proton collisions should be taken into account
on a scale of gamma-ray emitting region of GRBs.

\section{Summary}
\label{summary}

The Weibel instability in hot and dense counter-streaming relativistic plasma flows is considered
theoretically and numerically. The results include the following.

\begin{enumerate}[(i)]
    \item Due to relativistic pinch of angles, if the flows Lorentz factor $\Gamma \gg 1$, the
        instability scenario for hot plasma is the same as for cold one, namely current filaments
        elongated in the direction of the flows velocity, and the magnetic field focusing the
        filaments, are formed.
    \item At the linear stage of the instability transverse filament scale $\lambda_m$ and the
        instability growth rate $\Lambda_m$ can be estimated using Eqs.~\eqref{lambda_m} and
        \eqref{Lambda_m}. For certain Lorentz factor of the flows $\Gamma$ and proper flows
        temperature ($\propto \eta$) one can find $\Lambda_m \propto (\eta \Gamma)^{-1/2}$ and
        $\lambda_m \propto (\eta / \Gamma)^{1/2}$.
    \item \label{item_emission} Numerical simulations reveal that the generated magnetic field causes an efficient
        synchrotron emission by electrons, and the overall energy of the synchrotron photons can be
        much higher than the energy of the magnetic field.
    \item The criterion for judging copious pair production in Weibel instability is proposed (see
        Eqs.~\eqref{criterion} and \eqref{criterion_conclusion}). Moreover, fulfillment of this
        criterion also ensures that the energy of synchrotron photons is greater than the magnetic field
        energy~(\ref{item_emission}).
    \item The considered effects become noticeable for plasma with very high value of the mean electron
        Lorentz factor, that leads to the timescale of collisional effects much larger than the
        instability timescale.
    \item In the framework of the collapsar model of long gamma-ray bursts, $\varkappa \gtrsim 1$ and
        even $\varkappa \gg 1$ can be reached for the interaction of the jet with the progenitor
        star envelope, or for internal shock in the jet at the distance about $100$ Schwarzschild
        radii from the black hole (see Fig.~\ref{gammaN}).
\end{enumerate}

The Weibel instability that leads to $\varkappa \gg 1$ should potentially modify the plasma parameters
dramatically. The gamma-ray emission and the photon $e^+e^-$ pair production would not stop until the mean particle energy
becomes so low that $\varkappa \lesssim 1$. Therefore, in the shock region the plasma density can
be increased much due to the pair production, that at the same time leads to the decrease of the mean
particle energy. The impact of
this scenario on the GRBs models would be
considered elsewhere.

\section{Acknowledgements}

This research was supported by the Russian Foundation for Basic Research (Grant No.~15-02-06079), by the
Grants Council under the President of the Russian Federation (Grant No. MK-2218.2017.2) and by “Basis”
Foundation (Grant No.~17-11-101).

We thank Vl.~V.~Kocharovsky for inspiring conversations and I.~I.~Artemenko for discussion of
the effect of collisions.

\bibliography{main}
\end{document}